\documentstyle[12pt]{article}
\def\beq{\begin{equation}}
\def\eeq{\end{equation}}
\def\ben{\begin{eqnarray}}
\def\een{\end{eqnarray}}
\def\bea{\begin{array}}
\def\eea{\end{array}}

\begin{document} 

\baselineskip=20pt

\begin{flushright}
Prairie View A \& M, HEP-8-95\\

\end{flushright}

\vskip.15in

\begin{center}
{\Large\bf Small Elements in Fermion Mass Matrices and\\
 Anomalous Dipole Moments}

\vskip .5in

{\bf Dan-di Wu}\footnote{ $~$Electronic address:  Wu@hp75.pvamu.edu, or
DanWu@
physics.rice.edu}

{\sl HEP, Prairie View A\&M University, Prairie View, TX 77446-0355, USA}

\vskip .3in

\end{center}
\vskip1.in

\begin{abstract}
 
Assuming the small entries in the mass matrices are produced by 
fermion-scalar  loops,
we calculate the anomalous dipole moments of the leptons and
quarks. The top quark appears in all the loops as the mass seed.  
When comparing the results with experimental data, including electric
and magnetic dipole moments, and radiative transition rates, we obtain
the mass limits which are typically
 larger than .1 TeV for the relevant  neutral scalars,
 and 70 TeV for the relevant lepto-quarks. We then discuss 
the $P-\bar P$ mixing with a toy model. Rates of the known mixings
  require the masses of some neutral scalars to be 
large.
\end{abstract}

\vskip1.in
\begin{center}
\end{center}
\clearpage

An understanding of the masses and mixing of quarks and leptons (the fermions)
 is one of the 
 challenges of high energy physics. Thirty years ago, many studies 
followed an understanding of the spectrum of hadrons. Hopefully, the 
study of masses and mixing of fermions  will similarly open a new world
of physics study. Indeed, much effort has been made to solve this puzzle
in the past twenty years [1,2,3,4].
There are strong indications, from the previous results, that the mass problem
is deeply related to physics at very high scales, the grand unification 
scale, or 
even the superstring scale\footnote{For example, the minimal grand 
unification
theory predicts $m_b=m_\tau$ [2]. For recent studies in this direction, 
see [4].}.
\vskip.1in 

However the possibility of explaining at least 
part of the mass matrices by low energy 
physics is still very attractive. 
As an effort to find a window to  
a low scale explanation, we
 try here an  approach in which   small
entries in the mass matrices are assumed to be 
products of radiative corrections. 
 In other 
words,  the 
famous Higgs mechanism $m_f=Gv$ (where $G$ is the relevant Yukawa coupling 
constant and $v$ is the vacuum expectation value of the relevant Higgs field) 
is no longer responsible for the full texture of the mass matrices. 
There are some initial texture zeros (ITZs), 
which according to the previously suggested textures should 
be non-zeros. Early suggestions of ITZs can be found in [5]. 
Recently, the author suggested a different pattern  of ITZs [6], in which the
needed corrections to the ITZs are all less than .001$m_t$, 
which are comfortable for radiative corrections. Inversely, radiative
correction 
provides a new mechanism for an extra mass hierarchy. 
\vskip.1in

Once the bold assumption is made, the mechanism of mass hierarchy must play 
an active role in low energy phenomenology. The easiest low energy physics
to consider is the  anomalous dipole 
moments corresponding to the radiatively produced mass. The relation between 
the two important quantities is  controlled roughly by only one parameter, 
the largest mass of the involved scalars.  By confronting the
resultant dipole moments with  experimentation, the scale of new physics,
the physics of new scalars, is obtained.  
This study can be done model independently as shown below. One may go
along the model-independent way further (although this is not done in 
this paper).
However, in order to   simplify the relation of  this mechanism 
to other aspects of physics,    
we will discuss a toy model in the later stage (Section {\bf IV})
of this work. The  relevant loop diagrams in this model
are  obviously convergent. 
\vskip.1in

\vskip.1in
\noindent{\large\bf I. General Considerations}
\vskip.1in

A typical Feynman diagram for a radiatively produced mass matrix element
is shown in Fig. 1. Note that since this matrix element, whatever it is, 
vanishes at the tree level, the loop diagram
which produces this element must be convergent, 
for the sake of the renormalizability of the assumed 
theory. Therefore the boson line in 
this loop diagram must be made of two different bosons which mix with 
each other.
If Fig. 1 exists in the assumed theory and is responsible for the 
corresponding mass 
matrix, then the Feynman diagrams in Fig. 2  must exist too, where Fig. 
2$a$ is
only for neutral current couplings and Fig. 2$b$ should be added if charged 
couplings are involved. A comparison of the two diagrams leads  immediately to 
a relation between a mass matrix element and the corresponding dipole 
matrix element.  This relation actually sets a limit to the masses of the 
internal 
bosons, when combined with the relevant experimental data, because the most 
sensitive parameter in this relation is the heaviest mass in the loop.

\vskip.1in

As is well known, the loop diagram is proportional 
to the mass of the internal fermion $M$, so called the
mass seed, with a suppression factor $\kappa={1
\over 16\pi^2}\frac{\delta\mu^2}{\mu^2}g_1g_2$ where $\mu^2$ is 
the square of the largest mass
among the three masses in the loop. $\delta\mu^2$ is the mixing mass 
between the two
bosons, with $\frac{\delta\mu^2}{\mu^2} \le \frac{1}{2}.$ This suppression 
factor immediately excludes 
large matrix elements to be considered as radiatively produced, if new 
extremely heavy 
fermions are not introduced to provide the  mass seed $M$. The reason 
is simple
 because the largest fermion mass  in the standard model (SM) is  
the mass of the top quark $m_t$. Of course, it is still possible that 
some hidden heavy fermion sectors may provide the large  mass seed $M$. 
Then all 
the fermion mass matrix elements can be radiatively produced. However, 
in this article we will examine the most interesting scenario whereby the 
top mass is 
the mass seed for radiative production of small masses, 
although some of our conclusions may   
also be valid in other situations. This scenario is interesting because 
it is most
restrictive. 
\vskip.1in

It is unlikely that the  bosons in the loop 
are gauge bosons. It has been noticed  that if the gauge bosons are 
coupled to the  fermions with the same chirality, then Fig. 1 will be 
proportional to the larger
 of the outside fermion masses, which is too small. Even if the two gauge 
bosons
are coupled to fermions with different chiralities, the factor $g_1g_2$ 
will already be at 
 order  0.1. This suppression will further limit the usage of 
Fig. 1\footnote
{If the hidden heavy fermion sector is the so-called right handed mirror
fermions [7], then this mechanism may work, although the model is less 
phenomenologically restrictive.}.
In contrast,  the discovery  of the heavy top quark implies that
the corresponding Yukawa coupling constant is about 1. This provides 
quite a lot of room for the mechanism of radiative production of 
masses. 
\vskip.1in

Another line of thinking  also leads to the idea of radiative production 
of  part of the fermion mass matrices. Since the observation of the 
fermion mass hierarchy, many authors have speculated that perhaps the masses 
of the third family are produced first, and the other small masses are 
 produced later by different mechanisms [8]. A possible candidate of the
 second mechanism 
is radiative production[9]. Indeed, considering the order of magnitude of the matrix 
elements,
the following patterns for initial mass matrices are recommended [6]
\ben
M^F_i=\left(\bea{ccc}
0&0&0\\
	             0&0&-b^F\\
	              0&b^F &c^F\eea\right), 
\een
with $F= U,\ D,\ L$ for up, down and leptonic mass matrices,
where $c^U= m_t$, $c^D= m_b$ and $c^L= m_\tau$; and 
$c^U:b^U\sim c^D:b^D \sim \sqrt{m_t/m_c}$, 
$c^L:b^L\sim \sqrt{m_\tau/m_\mu}$. The small step hierarchical chain 
$c^U\rightarrow b^U\rightarrow c^D\rightarrow b^D\rightarrow
 c^L\rightarrow b^L$ is realized by a combination of sequentially smaller
 Yukawa coupling constants and vacuum expectation values.
Since there is not a principle to require the equalness of Yukawa couplings
 or VEVS in case of a multi-Higgs contribution to masses, a reasonable small 
difference cannot be ruled out. In Ref. 6 the (ambiguous) naturalness principle 
is appealed to explain the sequence of the Yukawa couplings.
In other words it was assumed that a smaller Yukawa coupling is originated
 from a smaller symmetry of the corresponding Yukawa term in the 
Lagrangian. It was found that the same philosophy cannot go beyond 
the initial pattern of Eq(1). Therefore a second mechanism
for mass hierarchy must be introduced in order to produce the whole mass matrix. By separating the initial elements from the initial texture 
 zeros (ITZs) we can at least
expect an extra hierarchy due to the 
loop suppression factor $\kappa$. Note, to be different, our initial
texture is not Hermitian.
The intended mass matrices produced by radiative corrections are then
\ben
M^F_r=\left(\bea{ccc}
0&-x^F&0\\
	              x^F&y^F&0\\
	              0&0 &0\eea\right),
\een
where $ y^D \sim m_s$, $y^U=y^L=0$, etc.  The scalars which are involved 
in the leptonic
loop diagram to produce the element $x^L$ from the top mass must be 
lepto-quarks which carry both the lepton number and the baryon number. Their 
electric charges could be either -1/3 (if the fermion in the loop 
 is the anti-top quark)
or -5/3 (if it is the top quark in the loop)
The total mass matrices for each 
type of the fermions are the sums of the corresponding matrices 
$$M^F=M^F_i+M^F_r$$
which is similar to one of the desired texture patterns suggested 
in the literature [10]. 
\vskip.1in

The corresponding dipole moment matrices are
\ben
 \mu^F=\left(\bea{ccc}
0&-x^{'F}&0\\
	             x^{'F}&y^{'F}&0\\
	              0&0 &0\eea\right).
\een
Note, in principle, the anomalous dipole matrices need not correspond to the
radiative mass matrices, Eq. (2). In particular, for instance, the (3, 3) 
element can be non-zero. However, its value is unrelated because of 
uncertainties in the internal parameters. Therefore,
we would rather make it vanishingly small. 
When the mass matrices are diagonalized by two unitary matrices\footnote{
Our symbols in matrices (1) to (3) are intended  just to show the pattern. 
In particular, we do not imply that the matrix elements are real. 
}, the dipole matrices should be
subject to the same rotation to become the dipole matrices 
in the mass representation,
\beq
\mu^{m}_F=U_L^F\mu^F U_R^F, \ \ \   
\eeq
 where $ U_L, \ U_R $ are
U-matrices such that   
$$  U_L\ M U_R $$ 
 is diagonalized with all eigenvalues positive.

\vskip.1in
\noindent{\large\bf II. The Results of the Loop Diagram Calculation}
\vskip.1in

The diagrams in Figs. 1 and 2 are all convergent and are solidly calculated.
For a radiative mass we have
\beq
m_{12}=\frac{M^*}{16\pi^2}\frac{\delta\mu^2}{\mu^2}g_1g_2C
\eeq
with $m=|M|$ and 
\beq
C=\mu^2\left\{{m^2\over (m^2-\mu_1^2)(m^2-\mu_2^2)}ln\ m^2
+{\mu^2_2\over (\mu^2_2-m^2)(\mu^2_2-\mu_1^2)} ln\ \mu^2_2
+{\mu_1^2\over (\mu_1^2-m^2)(\mu_1^2-\mu_2^2)} ln\ \mu_1^2\right\},
\eeq
where $\mu$ is the largest among $m\equiv m_t$, $\mu_1$, and $\mu_2$.
The value of $C$ is at  order  1. When one of the three masses 
is negligibly small, $C$ is convergent even if this small
mass is set to zero. For example, when $m=0$, we have [9]
$$C= \frac{\mu^2}{\mu_2^2-\mu_1^2}\ ln\ \frac{\mu_2^2}{\mu_1^2}.$$
However, setting any two masses vanishing will cause $C$ to be infrared 
(logarithmically)
divergent. One can, if one likes,  express $m_{12}$ in terms of mass
 eigenstates
of the relevant Higgs particles, instead of the eigenstates of 
interactions as in
Eq. (5). In this case, Fig. 1 will be replaced by two divergent diagrams,
 with 
their coupling constants satisfying a GIM [11,12] like unitarity 
condition which guarantees
the convergence of the sum of the two diagrams\footnote{The author thanks
  James Liu for 
providing such a discussion.}.  The magnitude of  $m_{12}$ is proportional 
to the
mass of the fermion in the loop, $m=|M|$. The hierarchy property of the
 fermion mass spectrum immediately leads to the dominance of the loop
diagram whose internal fermion is a top quark. Loops with lighter internal 
fermions can be completely ignored. This will greatly 
simplify the calculation and the analysis. 
It is worth noting that $m_{12}$ does not 
decrease with the scale
of the loop $\mu^2$. Instead, it is proportional to the 
ratio $\delta\mu^2/\mu^2$. This means somehow the radiatively
 produced mass 
matrix elements are immune from the ``decoupling theorem", which is typical
for physics related to Higgs particles [13]. Since too many parameters
appear in (5), other data are needed to extract a
specific piece of information. It turns out that the corresponding 
dipole moment is the most convenient of these.

\vskip.1in

For the corresponding anomalous dipole moments, a universal formula can
 be written as
\beq
\mu_{12}=e\frac{m_{12}}{\mu^2}\left[{Q_tC_t+Q_HC_H\over C}\right],
\eeq
where $Q_t+Q_H=Q_{\rm out}$. The second term in the brace appears only
when the electric charge of the Higgs is non-zero, $Q_H\ne 0$.
$\mu$ has been explained before. $C_t$ and $C_H$ are, 
respectively, 
\beq
C_t=\frac{\mu^4}{\mu_2^2-\mu_1^2}\left[\frac{\mu_2^2}{(m^2-\mu_2^2)^2}
ln\ \frac{m^2}{\mu^2_2}
-\frac{\mu^2_1}{(m^2-\mu_1^2)^2} ln\ \frac{m^2}{\mu_1^2}\right]-
\frac{\mu^4}{(m^2-\mu_1^2)(m^2-\mu_2^2)},
\eeq
\beq
C_H= \frac{\mu^4}{2\mu_1^2\mu_2^2}
\eeq
For $C_H$, we have only calculated  a simple case in which 
 $m^2 <<  \mu_1^2,\ \mu_2^2$. When masses of the two charged scalars are 
very different,  the $C_H$ term in (7) will dominate and the dipole moment
will have no longer a $1/\mu^2$ suppression.
Now, from Eq. (7) we see that we can indeed
extract some specific information about the 
loop scale once $m_{12}$ is somehow known.
The uncertainty factor in the squared bracket is a slow varying 
real function  of the
ratios of masses in the relevant loops (except when
$C_H$ term dominates). $\mu_{12} $ and $m_{12}$ share the 
same phase up to modular $\pi$, $\arg(\mu_{12})=\arg(m_{12}) + 
{\rm mod}[\pi]$ .
Furthermore this extraction can be done  
without digging into the detail of the underlying
physics, especially its complicated Higgs sector. Since so far no scalar has
ever been found, one may suspect the existence of the scalar particles; on the 
other hand, the extreme uncertainty  leaves room for bold speculations.

\vskip.1in
\noindent{\large\bf III. The Anomalous Dipole Moments}
\vskip.1in

We will not go into the 
details of how the corresponding relevant parameters
fit into the scheme. 
What makes such work  difficult is the lack of information about
 the concrete form of 
the mass matrices. What we have measured so far, in the case of
weak interactions of the quarks, are the (diagonalized) mass values and
their weak mixing (the so called CKM mixing matrix [12], which is the 
product of 
two unitary matrices, $V(CKM)=U_L^{U}U^{D\dagger}_L$). To skip these 
complications, we therefore assume 
that the following mass matrix is reached [10] as the sum of initial and 
radiative textures
\ben
M=\left(\bea{ccc}
0&-x&0\\
	             x&y&-b\\
	              0&b &c\eea\right).
\een
The diagonalized unitary matrices
are close to identity matrices (therefore they are called small 
unitary matrices), 
up to a diagonal phase matrix $P$. 
Without loss of generality we choose $U_L$ and $U'$ to be small 
unitary matrices and let $U_R=U'\ P$. These small unitary matrices are
typically of the form 
\ben
\left(\bea{ccc}
1\ \ \ \ \ &-\epsilon_1&-\epsilon_3+\epsilon_1\epsilon_2\\\
\epsilon_1^*\ \  \ \ &1&-\epsilon_2\\
\epsilon^*_3\ \ \ \ &\epsilon^*_2&1\eea\right)
\een
\vskip.1in

Applying this form to $MM^\dagger $ to obtain $U_L$ and to $M^\dagger M$ to
 obtain $U'$, we  obtain, to the leading orders of hierarchical quantities, 
for the left handed unitary matrices: 
\beq
\epsilon_{1L}= \frac{x}{y+\epsilon_{2L}b},\ \ \epsilon_{2L}= -\frac{b}{c},\ \ 
\epsilon_{3_L}=  0, 
\eeq
and the masses are
\beq
m_1=|\epsilon_{1L}x|,\ \ m_2= |y+\epsilon_{2L}b|,\ \ m_3=|c|.
\eeq
The parameters in $U' $ are 
\beq
\epsilon'_1=-\epsilon_{1L}^*,\ \ \epsilon'_2=-\epsilon_{2L}^*,\ \ 
 \epsilon'_3=0.
\eeq
The phase matrix $P$ is
\beq  
 P={\rm diag}(e^{-i(2\phi_1-\phi_2)},\ e^{-i\phi_2},\ e^{-i\phi_3})\equiv
(P_1,\ P_2,\ P_3),
\eeq
with $ \phi_1=\arg(x), \ \phi_2=\arg(y+b^2/c),\ \phi_3=\arg(c).$
These results are seen in the literature [10]
in different contexts with some changes. (Mainly because our mass matrices
are not Hermitian, in contrast to most of the literature which 
are Hermitian.) Corresponding to the three mass matrices for the up-, down-, 
and charged lepton-type of fermions (We do not consider massive neutrinos
in this paper.), there are altogether six U-matrices for diagonalization 
of the respective mass matrices. All of them are physically relevant as will 
be shown below. Only one combination of the six U-matrices makes the CKM matrix. 
The other five different combinations represent physics beyond the CKM matrix.
\vskip.1in
  
Since the dipole matrices are not 
proportional to their corresponding mass matrices, miracle enhancement or 
suppression of the dipole moments which already exist in Eq. (3) 
is not expected after diagonalization, except  
that the zero elements in (3) may become non-zero. Indeed, the dipole moment
in the mass representation is
\ben
\mu^m
=\left(\bea{ccc}
2\epsilon_{1}x'P_1-\frac{m_1}{m_2}y'P_2&(-x'+\epsilon_{1}y')P_2
&(-\epsilon_{2}^*x'+\epsilon_{1}\epsilon_{2}^*y')P_3\\
(x'-\epsilon_{1}y')P_1&(2\epsilon_{1}^*x'+y')
P_2&(\epsilon_{1}^*\epsilon_{2}^*
x'+\epsilon_{2}^*y')P_3\\
(-\epsilon_{2}^*x'+\epsilon_{1}\epsilon_{2}^*y')P_1&
(-\epsilon_{1}^*\epsilon_{2}^*
x'-\epsilon_{2}^*y')P_2&
-\epsilon_{2}^{*2}y'P_3
\eea\right),
\een
where $\epsilon_i=\epsilon_{Li}$.
The calculation has been consistently done under leading order approximations, 
based on the hierarchical property of the mass matrices. Of course there are 
three kinds of such dipole matrix corresponding to flavor changed (and neutral)
 EM processes among U-type, D-type, and L-type fermions respectively.
\vskip.1in
 
These dipole matrices
need immediately to be compared with experiments. The most sensitive 
are the electrical dipole moments of the lepton and the neutron.
Note since $y'=0$ for 
leptons, the electrical dipole moment of the electron is
 zero, compatible with the 
experiments [15,16]. The electrical dipole moment of the
neutron comes from that of the 
u-quark ($D_n^u$) and that from the d-quark ($D^d_n$). We express this 
as $D_n
=D_n^u+D_n^d=\frac{4}{3}d_d-\frac{1}{3}d_u$, with $d_u,$ and $d_d$, the
E-dipoles of the u-quark and the d-quark respectively. Since $y'^U=0$, 
$d_u=0$.
 For the d-quark, 
\beq
 D^d_n\sim\frac{4}{3}e\frac{m_d}{\mu^2}\left(\frac{\frac{2}{3}
C_t-C_H}{C}\right)
{\rm arg}(m_s-\epsilon^D_2b^D).
\eeq
$\mu$, the mass  of one of the charged scalars  in the $y^D$ loop, has to 
be  20 TeV, in order to obtain the needed suppression [17] (the angle in the 
formula is taken  as 0.1.) 
 The experimental 
and theoretical uncertainties of the magnetic moments of the electron and the 
muon are both 
10$^{-22}$e$\cdot$cm [18,19]. These are comfortable with the  scale 1.4 TeV
 for the lepto-quarks in the relevant loop.
However, the process $\ \mu \rightarrow e  +\gamma\ $  puts a stronger 
limit to
the same lepto-quark mass. The results on dipole moments are summarized 
in Table 1. In this table, experimental data are compared with 
the minimal standard model predictions. The scales of the dipole loops 
(Fig. 2)
are obtained based on the experimental data. For the names of the scalars   
 which appear in
the table, see the next section. From this table we can see that the 
restrictions from
dipole moment experiments
to the masses of the neutral scalars, which couple to $t\bar c$ or $t\bar u$,
may not exist. Therefore, the possibility of 
 a top to on-shell scalar decay is not ruled out, because a light scalar 
may exist.
\vskip.1in

However, when we come to a specific model, this possibility needs to 
be reexamined.
In particular, we will be able to calculate the $P-\bar P$ mixing due to 
scalar mediated flavor changed neutral currents. The mixing mass here
will be proportional to $\mu^{-2}$ of the relevant scalar. It is also very 
sensitive to the relevant Yukawa couplings.
The Yukawa couplings for each of the scalars involved are  
originally given in the
interaction representation of the fermions. 
When we discuss physics in the 
mass representation of the fermions, we also need to transfer these 
couplings 
into the mass representation. This will be exemplified in the 
following section.

\vskip.1in
\noindent{\large\bf IV. Physics with  an ITZ Toy Model}
\vskip.1in
 
Our previous discussion on the dipole moments is model independent. However, 
we would like  to  proceed further, in particular to 
understand the implications of the obtained mass limits in Table 1 
and compare 
these limits with the limits obtained elsewhere. 
To work with a specific model will make such a discussion much easier, 
and much more specific. 
Therefore, we will introduce a toy model which will be able to
 produce the desired ITZs in (1) 
and the radiative corrections in (2). This model enjoys less symmetries in 
the Yukawa terms compared with that in Ref. [6]. However this model provides 
less stringent limits for the masses of the relevant scalars.
This by no means that the limits obtained here are the lowest possible. On
the other hand, one may hold a different philosophy so to keep the terms as
 symmetric as possible. Then the mass scales of the new scalars will 
be much higher, most of them in the beginning of the desert of GUT.
\vskip.1in

The gauge sector and the fermion gauge interaction sector of this model 
is completely 
standard, however we have a  complicated Higgs sector and 
a complicated Yukawa sector. 
Let us plainly write down the whole Yukawa interactions, since our main 
concern here
is the Yukawa sector:

\ben
\bea{ccc}
{\it \large L}_Y
&=&G_1\left[\bar\psi_L^3\Phi_{33}U_R^3+
\sqrt{\frac{1}{2}}\bar\psi_L^2\Phi_{23}U_R^3
+\sqrt{\frac{1}{2}}\bar\psi_L^1\Phi_{13}U_R^3+
\sqrt{\frac{1}{2}}\bar\psi_L^3\Phi_{23}U_R^2\right]+ h.c.\\
&&\\
&+&G_2\left[\bar\psi_L^3\xi^1U^2_R -\bar\psi_L^2\xi^1U^3_R -
\bar\psi_L^3\xi^2U^1_R\right] \hskip1.6in+h.c. \\
&&\\
&+&G_3\left[\bar\psi_L^3\phi^{33}D_R^3+\sqrt{\frac{1}{2}}
\bar\psi_L^3\phi^{23}D_R^2
\right]\hskip2.in+ h.c\\
&&\\
&+&G_4\left[\bar\psi_L^3\eta_1D^2_R-\bar\psi_L^3\eta_2D^1_R
-\bar\psi_L^2\eta_1D^3_R\right]\hskip1.5in + h.c.\\
&&\\
&+& {\rm leptonic\ part}\hskip.3in+\hskip.3in  {\rm lepto-quark\ part}.\hskip.3in\\
\eea
\een
We may identify $\Phi_{ij}$ (and $\phi^{ij}$) ($i,\ j=1,\ 2,\ 3$) 
as a set of $SU(3)$ sextet (and anti-sextet).  However, 
part of their components is missing in the Yukawa sector.  
$\eta_i$ (and  $\xi^i$) are triplet (and anti-triplet) Higgs fields 
in the same sense.  These Higgs fields are also all  $SU(2)_L$ doublets.
The $SU(3)$ here is a global 
symmetry group. All left handed fermions are in triplet (3) of 
$SU(3)$ and right-handed fermions are in $3^*$. There is also a global 
$U(1)$ symmetry. 
The $U(1)$ charges follow the formula $\xi=I-L/3$, where $I$ is the 
$SU(3)$ index of the multiplet. For the Higgs potential and how Higgs 
develop VEVs, see Ref. [20].
\vskip.1in
The ratios of the Yukawa coupling constants are assumed to be 
\beq 
G_1:G_2 \sim 3,\ \   G_2:G_3\sim 2, \ \ \ G_3:G_4\sim 5.    
\eeq
Therefore the Yukawa coupling constants in this model are all compatible.
Such small differences of Yukawa couplings will, combined with the 
differences in VEVs of the four Higgs fields $\Phi_{33}$, $\xi^1$, 
$\phi^{33}$, $\eta_1$, produce the needed small step hierarchy in the 
initial pattern of the mass matrices as that in (1). 
Therefore after SSB we will have the following
initial mass matrices:
\ben
M^U_i=\left(\bea{ccc}
0&0&0\\
	             0&0&G_2V'\\
	              0&-G_2V'&G_1V\eea\right),
\een

\ben
M^D_i=\left(\bea{ccc}
0&0&0\\
	              0&0&G_4v'\\
	              0&-G_4v' &G_3v\eea\right),
\een
\ben
M^L_i=\left(\bea{ccc}
0&0&0\\
	              0&0&G'_4v'\\
	              0&-G'_4v' &G'_4v\eea\right),
\een
There are five initial texture zeros (ITZs) in each mass matrix.
We assume
$$V : V',\sim 3.6, \ \ V':v \sim 1.7,\ \ v:v' \sim 2.1.$$ 
Of course, $\sum {\rm VEV^2}=245^2 {\rm GeV^2}$. 
Note that the initial mass hierarchy in this toy model is a combined 
effect of sequentially smaller coupling constants and the 
 VEVs. All ITZs in these matrices are protected by the symmetry 
property of the Yukawa sector. Actually all naive 
corrections to ITZs vanish, unless Higgs mixing is introduced, as is 
in Fig. 1. 
\vskip.1in

Eq. (5) can be applied to the specific Yukawa interactions,
with $\mu_1,\ \mu_2$ masses of the specific Higgs particles which appear 
in the
specific diagrams, and $\delta\mu^2$ their respective mixing
masses squared. We assume that the factor $g_1g_2M^*$ in (5) are 
replaced respectively by 
the specific coupling constants in the model as in 
the following 
\beq
 x^U\propto |G_1|^2G_2V^*/\sqrt{2},\ \ 
y^D\propto |G_1|^2G_3V^*/2,\ \ x^D\propto |G_1|^2G_4V^*/\sqrt{2},\ \
x^L\propto \lambda_1\lambda_2G_1^*V^{*}
. 
\eeq
One can easily recognize, for instance, in order to produce the (2,1) element 
$x^U$, the mixing between
$\Phi^0_{23}$ and $\xi^{2\ 0}$ must be introduced, and to produce the 
(1,2) element,
that between $\Phi^0_{13}$ and $\xi^{1\ 0}$ must be introduced. 
In general, these 
two elements may be different, however, for simplicity, we will only analyze
the case when they follow  the pattern in (2). A general analysis is
 not difficult 
to work out.
Once these small elements are obtained, all the analyses summarized in 
Table 1 will follow. It is worth pointing out that
 electric dipole moment of the d-quark is generally nonzero in this model.
Indeed, the angle in (17) is decided by the phase of
$$\frac{{(b^{D})}^2}{c^D}\frac{1}{y'^{D}}\propto \frac{(G_4v')^2}{G_3v}
\frac{1}{G_3V^*}.$$
This quantity  is rephasing invariant, unless there is an exact symmetry 
which
 allows separate changes of the phases of different vacuum expectation
values.
\vskip.1in
 
Some exotic processes are allowed in this model, if suitable scalar mixing is 
introduced. For example, $t \rightarrow c\ (u) s\bar b$ is
allowed if the mixing between the neutral
components   $\Phi_{23}^{ 0}\ (\Phi_{13}^{0})$
and $\phi^{23\ 0}\ $ exists; and $t \rightarrow c\  b\bar b$
exists if $\Phi_{23}^{0}-\phi^{33\ 0}$ mixing exists etc.
 However since these mixings 
are not used in our small
mass matrix element calculations, the magnitude of mixing here
can be any small, therefore the width of these processes can be any small,
unless $\Phi_{23}^{0}\ (\Phi_{13}^{0})$ is lighter than the top quark. 
The possibility of a light
$\Phi_{23}^{0}\ (\Phi_{13}^{0})$ is not ruled out by the dipole experiments.
However, in this specific model, it could be ruled out because it 
also mediates
 flavor changed neutral currents which may cause the $K-\bar K$ like mixing.
Existing data on the mixing of such systems are very stringent.
But before going into a detailed analysis, let us first find  out the 
Yukawa couplings in the mass representation. 
\vskip.1in

The quark neutral current 
Yukawa interactions in (18) are typically of the pattern
\ben
\left(\bea{ccc}
0\ \ \ & 0\ \ \ &\ \ \ a\\
0\ \ \ &0\ \ \  &\ \ \ b\\
a'\ \ &b'\ \ &\ \ \ c\eea\right).
\een
When the quarks are transformed into the eigenstates of masses, the 
corresponding Yukawa coupling becomes (The phase factor $P$ is 
neglected here.)
\ben
\left(\bea{ccc}
\epsilon[a+a'+\epsilon_1\Delta b+\epsilon c]\hskip.35in
&-\epsilon_2[a-|\epsilon_1|^2a'+\epsilon_1\Delta b+\epsilon c]&
a+\epsilon_1(b+|\epsilon_2|^2b')+\epsilon c\\
\epsilon_2[a'+|\epsilon_1|^2a+\epsilon_1\Delta b+\epsilon c]
&\epsilon_2[\epsilon_1^*(a'-a)
-\Delta b-\epsilon_2c]\hskip.25in
&-\epsilon^*_1a+b+|\epsilon_2|^2b'+\epsilon_2c\\
a'-\epsilon_1b'-\epsilon_1|\epsilon_2|^2b+\epsilon c\hskip.15in
&\epsilon_1^*a'+b'+|\epsilon_2|^2b-\epsilon_2c\hskip.3in&
-\epsilon_2^*\Delta b+c\hskip.75in\eea\right),
\een
where $\epsilon=\epsilon_1\epsilon_2, \Delta b= b-b'$.
Applying this form to $G_1$ couplings, we find that, for example, 
the contribution of $\Phi^{ 0}_{23}$  to $D-\bar D$ mixing vanishes. This is
an accidental fact of this specific model. Consequently, the mass of 
$\Phi^{ 0}_{23}$ is not limited by any existing $P-\bar P$ mixing data.
However the contribution of $\Phi^{ 0}_{13}$ to $D-\bar D$ mixing is
proportional to $(\epsilon_1^U\epsilon_2^U)^2$,
 which is not  big enough 
as a suppression. Therefore the mass  of $\Phi^{ 0}_{13}$ needs to be larger 
than 2.1 TeV. Such a heavy $\Phi^{ 0}_{13}$ is an indication that the 
$SU(3)$ symmetry is explicitly badly broken also in the Higgs sector, because
otherwise this particle would be a light pseudo-Nambu-Goldstone particle.
The mass limits for other neutral scalars from similar 
considerations are summarized in Table 2. We can see that in this 
specific model, 
the bounds obtained from the $B_d-\bar B_d$ mixing are all overwritten
by those obtained from the $K -\bar K$ mixing. 
 Comparing Table 1 and Table 2, we still find  attractive 
scenarios for the decay of top to an on shell neutral scalar\footnote
{Besides the simple
$t \rightarrow c + \Phi^{ 0}_{23}$ mode, 
such a decay through a mixing between $\Phi^{ 0}_{23}$ (in case it is 
too heavy to
be on shell in, for example, some alternative models) and a light scalar 
is discussed in [6].}, $\ \ t \rightarrow c + \Phi^{ 0}_{23}$. 
The life time of $\Phi_{23}^0$
will be relatively long because both of the smallness of the relevant mixing 
and of the  heaviness of $\phi^{23\ 0}$ ($>$4.2 TeV, if it decays into 
$\bar b s$) and $\phi^{33\ 0}$ 
($>$ .53 TeV, if it decays into $\bar b b$).

\vskip.1in
\noindent{\large\bf V. Concluding Remarks}
\vskip.1in

The puzzle of the pattern of masses and mixing of quarks and leptons
is believed to be deeply 
related to physics both at high scales and low scales.
Here we have made an effort to connect the mass matrices of the 
fermion with other low energy physics phenomena. Our approach is based on an
almost model-independent relation between a radiatively produced mass and 
an anomalous dipole moment, and an assumption on the initial texture zeros
in the mass matrices. 
\vskip.1in

By doing so we find that if the desired small elements in the fermion mass 
matrices are   radiatively 
produced  from a top mass seed, then masses of the new scalar 
bosons which are needed for this mechanism to work are above .1 TeV
for neutral scalars and up to 70 TeV for charged scalars,
in order to fit the known data on  electric and magnetic dipoles of 
the fermions. 
The highest scale for this mechanism to work is two orders of magnitude higher
than the weak scale, however it is well below the scale of grand unification.
Our study with a toy model in section {\bf IV} has enhanced the scale of 
some neutral scalars up to 4 TeV, although the model, 
in particular its parameters, are subject to 
optimization.
\vskip.1in

Since the scale of concern in this approach is low, other low-energy physical
processes should also be examined, except for the dipoles and the $P-\bar P$ 
mixing discussed here. Exotic processes, especially exotic top decays,  which do not exist in the minimal 
standard
model are generally expected in this approach. In addition, more
accurate data, 
such as $B_r(D\rightarrow \rho\gamma)$, will help further clarify the 
scale of
radiative corrections. The improved measurement of $\Delta m$ for the 
$D^0-\bar D^0$
system will be crucial for the scale of the toy model discussed in section 
{\bf IV}. Since we allow some scalar bosons to be as light as .1 TeV, their 
effects in some other loops will be also measurable and are therefore worth 
calculating.
\vskip.1in

The author is indebted to Dr. Zhi-zhong Xing for numerious discussions and
conversations over  
email which constantly stimulated this work. Part of this work was completed 
while the author was visiting 
Texas A\&M University. He thanks Professor R. Arnowitts for his 
hospitality. Discussions 
with  Dr. J. Liu at Texas A\&M is deeply appreciated. 
This work is supported in part by a grant from the National 
Science Foundation,
HRD-9550702.
\vskip.1in
{\bf Note added in proof}: I thank Dr. A.L. Kagan for bringing my attention to
his paper[9] which obtained $\mu_{12}\propto m_{12}$, similar to (7) here.

\newpage
\noindent{\large\bf References:}
\begin{itemize}
\begin{enumerate}
\item
S. Weinberg, Transaction of the New York Academy of Sciences, 
Series II, Vol. 38, 
185(1977);  H. Fritzsch, Phys. Lett. B70 (1977) 436; Phys. Lett. B73 
(1978) 317.
 \item 
H. Georgi and S.L. Glashow, Phys. Rev. Lett. 32 (1974) 438; A. De Rujula,
 H. Georgi, and
S.L. Glashow, Ann. Phys. (N.Y.) 109 (1978) 258;
M.S. Chanowitz, J. Ellis and M.K. Gaillard, Nucl. Phys. B128 (1977) 506; 
A. Buras, J. Ellis, M.K. Gaillard and D.V. Nanopoulos, Nucl. Phys. B135 
(1977) 66;
F. Wilczek and A. Zee, Phys. Lett. B70 (1977) 418.
\item  See, e.g. S. Pakavasa and H. Sugawara, Phys. Lett. 73B, 61 (1978);
R. Barbieri, R. Gatto, and F. Strocchi, Phys. Lett. 74B (1978) 344; 
T. Kitazoe and K. Tanaka, Phys. Rev. D18 (1978) 3476; D. Wyler, 
Phys. Rev. D19 (1979) 330
 D.D. Wu,
Phys. Lett. 85B (1979) 364. F. Wilczek and A. Zee, Phys. Rev. 
Lett.42 (1979)421;
T. Maehara and T. Yanagida, Prog. Theo. Phys. 60(1978) 822; A. Davidson, 
M. Koca, and K. C. Wali, Phys. Rev. Lett. 43 (1979) 92; J. Chakrabarti, 
Phys. Rev. D20 (1979) 2411; C.L. Ong, Phys. Rev. D22 (1980) 2886.
\item
 P.H. Frampton and O.C. W. Kong, IFP-713-UNC; 
C. D. Froggat and H. B. Nielson, Nucl. Phys. B147 (1979) 277;
L. Ibanes and G.G. Ross, Phys. Lett. B332 (1994) 100;
H. Georgi and C. Jarlskog, Phys. Lett. B86 (1979) 297;
G. Anderson, S. Raby, S. Dimopoulos, and L. J. Hall, Phys. Rev. 
D47 (1992) 3702;
G. F. Giudice, Mod. Phys. Lett. A7 (1992) 2429;
G. K. Leontaris and  J. D. Vergados, Phys. Letts. B305 (1993) 242; 
S. Babu and Q. Shafi, BA-95-06 and Hep-ph/9503313;
S. Babu and R. N. Mahapatra, Phys. Rev. Lett.74 (1995) 2418; K.C. Chou 
and Y.L. Wu, CAS-HEP-T-95-06/001 
R. Peccei, in preparation.  
 \item 
See, e.g.,  
T. Terazawa, Phys. Rev. D22 (1980) 2921; D41 (1990) E3541; 
D. D. Wu, Phys. Rev. D34 (1986) 280;
 P. Kaus and S. Meshkov, Mod. Phys. Lett. A3 (1988) 1251.  
\item
D.D. Wu, in proceedings of Workshop on Particle Theory and Phenomenology
 May 17-19, 1995, Ames, Iowa, World Scientific, forthcoming, LALN
Bulletin HEP-PH/9510234;
D.D. Wu, Prairie View A\&M preprint, HEP-3-95.
\item
 J.C. Pati,  in Proceedings of Superstrings, Unified Theories and Cosmology
1987, World Scientific, p. 362.
\item
 See, Ref. [5], and   
S. Weinberg, UTTG-05-91, (Contribution to a volume in honor of Baqi Beg).
\item For using top mass as mass seed, see, X.G. He, R.R. Volkas, and D.D. Wu,
Phys. Rev. D41 (1990) 1630. For the radiative mass calculations, please compare
 with e.g.  P.H. Frampton, P.I. Krastev, and J.T. Liu, Mod. Phys. Lett.
A9 (1994) 761; 
S.M. Barr, Phys. Rev. D21 (1980) 1424. Also see, A.L. Kagan, Phys. Rev. D51 
(1995) 6196.
\item  
See Ref. [1] and 
D.S. Du and Z.Z. Xing,
Phys. Rev. D48 (1993) 2349; P. Ramond, R.G. Roberts, and G.G. Ross, Nucl. 
Phys. B406 (1993) 19; H. Fritzsch and Z.Z. Xing, Phys. Lett. B353 (1995) 114.
Z.Z. Xing, Private communications.
\item 
S.L. Glashow, J. Iliopoulos, and L. Maiani, Phys. Rev. D2 (1970) 1285. 
\item 
N. Cabibbo, Phys. Rev. Lett. 10 (1963) 531; M. Kobayashi, and T. Maskawa,
Prog. Theor. Phys. 49 (1973) 652.
\item 
T. Inami and C.S. Lim, Prog. Theo. Phys. 65 (1981) 297. 
\item
 See, for example,  D.S. Du and Z.Z. Xing, Ref. 10. 
\item 
K. Aboullah et al, Phys. Lett. Rev. 65 (1990) 2347; 
S.A. Murthy et al, Phys. Rev. Lett. 63 (1989) 935.
\item 
J. Bailey et al, Nucl. Phys. B150 (1979) 1.
\item
I.S. Altarev et al, Phys. Lett. B276 (1992) 242; K.P. Smith et al, 
Phys. Lett. B234 (1990) 191.
\item
E.R. Cohen and B.N. Taylor, Rev. Mod. Phys. 59 (1987) 1121; R. S. Van Dyck, et al,
 Phys. Rev. Lett. 59 (1987) 26.
\item
Cohen, Ref. 18.
\item
 H. Ruegg, Phys. Rev. D22 (1980) 2244; D. D. Wu, Nucl. Phys. 199B (1981) 523.
\item
R. D. Bolton et al, Phys. Rev. D38 (1988) 2077.
\item
CLEO Collaboration, R. Ammar, et al. Phys. Rev. Lett. 71 (1993) 674.
\item
See e.g. T. Tang, J.H. Liu and K.T. Chao, Phys. Rev. D51 (1995) 3501; 
P.A. Griffin, M. Masip, and M. McGuigan, Phys. Re.v D. 50 (1994) 5751;
 A. Kapustin, Z. Ligeti and H. D. Politzer, CALT-68-2009;
A. Ali and V.M. Braun, DESY 95-106; T. Inami and
 C.S. Lim, Ref. 13, M.K. Gaillard and B. Lee, Phys. Rev. D10 (1974) 897.
\item
CLEO Collaboration, M. Athanas et al. in proceedings of ICHEP94, Ref. GLS0159.
\item
CLEO Collaboration, P. Haas et al, Phys. Rev. Lett. 60 (1988) 1614.
\item
E691 Collaboration, J. C. Anjos et al, Phys. Rev. Lett. 60 (1988) 1239.
\item
 CDF Collaboration, F. Abe et al, Phys. Rev. Lett. 72 (1994) 3456; DELPHI 
Collaboration, P. Abreu et al, Z. Phys.
C57 (1993) 181;  OPAL Collaboration, Acton et al, Phys. Lett. B307 (1993) 247; 
ALEPH Collaboration, D. Buskulic et al, Phys. Lett. B307 (1993) 194.
\item
See, e. g. L.M. Barkov, Yad. Fiz, 46 (1987) 1088.
\end{enumerate}
\end{itemize}
 
\newpage 

\begin{center}
{\bf \large Table 1. Dipole Moments of the Fermions}
\end{center}
$$\bea{|c|ccc|}
\hline
&&&\\
{\rm dipole}\ &\ {\rm exper.} \ & \ \ \ {\rm MSM }\
\ \ & \ \ {\rm ITZ\ scalar\ mass\ ({\rm TeV}) } \ \ \\
&&&\\
\hline
&&&\\
d_e& <10^{-26}{\rm e}\cdot {\rm cm}[15]&  0  & {\rm no\ bound}\\
&&&\\
d_\mu&<10^{-19}{\rm e}\cdot {\rm cm}[16]& 0& {\rm no\ bound}\\
&&&\\
\Delta \mu_e\ \&\ \Delta\mu_\mu&<10^{-22}{\rm e}\cdot {\rm cm}[18,19]& {\rm compatible} &
\mu(lep-qrk)>4.5\ \\
&&&\\

D_n&<10^{-25}{\rm e}\cdot {\rm cm}[17]&10^{-31}{\rm e}\cdot {\rm cm} &  \ \ 
\mu_{\Phi_{23}^{ -  }},\ \mu_{\phi^{23\ +}}> 20\\
&&&\\
\mu\rightarrow e \gamma&Br<5\times 10^{-11}[21]&Br=0&\mu({\rm lep-qrk})>70\\
&&&\\
B\rightarrow K^*\gamma&Br=5\times 10^{-5}[22]&{\rm compatible}[23]&\mu_{\phi^{23\ +
}}> 0.25\\
&&&\\
B\rightarrow \rho\gamma&Br<2\times 10^{-5}[24]&{\rm compatible}&{\rm no\ bound}\\
&&&\\
D\rightarrow \rho\gamma &Br<10^{-1}[25]&Br<10^{-6}&{\rm no\ bound} \\
&&&\\
\hline
\eea
$$

\begin{center}
{\bf \large Table 2. Bounds from Data on $P-\bar P$  Mixing Systems}
\end{center}
$$\bea{|c|ccc|}
\hline
&&&\\
{\rm P-\bar P}\ &\ {\rm \Delta m_{exp}}\ ({\rm eV}) \ & \ \ \ {\rm MSM\ 
(eV)}\
\ \ & \ \ {\rm ITZ\ scalar\ mass\ (TeV) } \ \ \\
&&&\\
\hline
&&&\\
D-\bar D&<1.4\times 10^{-4}[26]&3\times 10^{-7}&\mu_{\Phi^0_{33}}> .27,\
\xi^{1\ 0}>2.0,\\
&&&\mu_{\Phi_{13}^0}>2.1,\ \xi^{2\ 0}>1.0 \\
&&&\\
B_d-\bar B_d& 3.5\times 10^{-4}[27]&{\rm compatible}&\mu_{\eta_1^0}>.42,\ \mu_{\phi^{23\ 0}}>.14\ \\
&&& \mu_{\phi^{33\ 0}}>.20\hskip.6in\ \\
&&&\\
K-\bar K& 3.5\times 10^{-6}[28]&{\rm compatible}&\mu_{\eta_1^0}>2.4,\ 
\mu_{\phi^{23\ 0}}>4.2\\
&&&\mu_{\eta_2^0}>1.2,\ \mu_{\phi^{33\ 0}}>.53
\\
&&&\\
B_s-\bar B_s& ?&0.007&\Delta m \sim .007\ {\rm eV}\\
&&&({\rm if}\ \mu_{\phi^{23\ 0}}\sim 4.2 ) \\
&&&\\
\hline
\eea
$$
\end{document}